\documentclass[preprint,12pt]{elsarticle}

\usepackage{cmap}
\usepackage{tikz}
\usepackage{pgfplots}
\usepackage{amssymb,amsmath,graphicx}
\usepackage[T1]{fontenc}
\usepackage{mathtools}
\usepackage{hyperref}
\usepackage{verbatim}
\usepackage{siunitx}
\usepackage{algorithm,algpseudocode}
\pgfplotsset{compat=newest}
\usepgfplotslibrary{fillbetween}
\usetikzlibrary{intersections}
\usetikzlibrary{patterns}
\pgfdeclarelayer{bg}
\pgfsetlayers{bg,main}
\usepackage[margin=1.25in]{geometry}

\newcommand{\xcap}{ {\bf  \hat{x} } }
\newcommand{\ycap}{ {\bf  \hat{y} } }
\newcommand{\zcap}{ {\bf  \hat{z} } }

\newcommand{\angstrom}{\text{\normalfont\AA}}

\journal{Journal Name}

\begin{document}

\begin{frontmatter}



\title{Solving Lubrication Problems at the Nanometer Scale}

\author{Nisha Chandramoorthy and Nicolas G. Hadjiconstantinou}

\address{Department of Mechanical Engineering\\ Massachusetts Institute of Technology\\ Cambridge, MA 02139\\ United States}

\begin{abstract}
Lubrication problems at lengthscales for which the traditional Navier-Stokes description fails can be solved using a modified Reynolds lubrication equation that is based on the following two observations: first, classical Reynolds equation failure at small lengthscales is a result of the failure of the Poiseuille flowrate closure (the Reynolds equation is derived from a statement of mass conservation, which is valid at all scales); second, averaging across the film thickness eliminates the need for a constitutive relation providing spatial resolution of flow profiles in this direction. In other words, the constitutive information required to extend the classical Reynolds lubrication equation to small lengthscales is limited to knowledge of the flowrate as a function of the gap height, which is significantly less complex than a general constitutive relation, and can be obtained by experiments and/or offline molecular simulations of pressure driven flow under fully developed conditions. The proposed methodology, which is an extension of the Generalized Lubrication Equation of Fukui \& Kaneko to dense fluids, is demonstrated and validated via comparison to Molecular Dynamics simulations of a model lubrication problem.  

\end{abstract}




\end{frontmatter}



\section{Introduction}
\label{intro}
Molecular Dynamics (MD) simulation \cite{tildesley,frenkel} has been a valuable asset in the study of nanoscale fluid mechanics and transport, 
finding extensive use  by researchers studying a variety of nanoscale problems, ranging from flows
around and through carbon-based nanostructured materials \cite{md3,md4,md5,md6} to boundary lubrication problems \cite{zheng,berro,savio}.
In some cases, the systems of interest are sufficiently small, that a direct simulation is feasible \cite{berro,zheng1}. In other cases, when direct simulation 
is not possible, MD has been used to provide insight into the physics of these flows \cite{landman,priezjev}. 

However, despite the insight gained on a number of aspects of nanoscale flow by MD studies, a general description of such flows beyond Navier-Stokes has yet to be developed for dense fluids. In fact, in many instances, different studies reach contradictory conclusions \cite{BeskokJChemPhys}, with the only clear consensus being that the Navier-Stokes description remains remarkably robust, at least up to scales as small as $O(10 \;{\rm nm})$. This can be qualitatively explained by noting that, for a dense fluid, the characteristic fluid lengthscale, $\hat{\sigma}$, is on the order of the molecular size; deviations from Navier-Stokes, expected as the characteristic flow lengthscale becomes smaller than $O( 10 \;\hat{\sigma})$ \cite{evans97,Liakopoulos2009,Beskok}, should therefore manifest themselves at lengthscales of $O(1-10 \;{\rm nm})$, depending on the fluid. A similar argument correctly predicts Navier-Stokes breakdown in rarefied gases to occur at much larger, $O(\mu {\rm m})$, scales \cite{nicolas}, owing to the significantly larger size of the mean free path compared to the molecular size.

 The present paper focuses on a particular class of problems, namely of the lubrication type, for which progress can be made using the following observation, originally due to Fukui and Kaneko \cite{fukui}:  in lubrication problems, averaging across the film thickness (made possible by the small film thickness) eliminates the need for spatial resolution of flow profiles in the transverse film direction. In other words, knowledge of the flowrate due the local pressure gradient is sufficient for closing the governing (lubrication) equation, thus reformulating the problem from one of finding the general constitutive relation for the stress tensor  to that of finding the {\it fully developed} flowrate in response to a pressure gradient. The importance of this simplification cannot be overstated: lubrication theory \cite{leal,szeri} separates the effects of axial variations (captured by the lubrication equation) from the constitutive relation (for the flowrate), allowing the latter to be determined once and for all in the significantly simpler, lower-dimensional setting of the fully developed flow (with no axial variation).
 
 As stated above, this approach has already been exploited by Fukui and Kaneko \cite{fukui} who have developed a "Generalized Lubrication Equation"  (GLE) for treating dilute-gas lubrication problems. In their case, they used solutions of the Boltzmann equation to develop a description of the pressure-driven flowrate. Their extended lubrication equation has been used extensively by other researchers to study a variety of problems of practical interest, ranging from air bearings \cite{garcia} to squeeze-film damping in electromechanical devices \cite{gallis} and has been reviewed extensively (for example, see \cite{cercig1,szeri}). Although the dense-fluid-lubrication research community has made significant strides towards treating nanoscale lubrication phenomena \cite{nanolub1,nanolub2,nanolub3} (as well as other complex physical phenomena, such as cavitation), it appears to have overlooked the GLE approach and its potential. Our objective here is to highlight this potential, demonstrate the feasibility of the approach using an example test problem and finally highlight some of the challenges and open problems unique to the dense fluid case.
 
 Here, we would be amiss to not discuss the work of Borg et al. \cite{borg}, who also proposed a general method for solving multiscale problems of high aspect ratio and would thus be applicable to the lubrication problems discussed here. Their approach is based on the Homogeneous Multiscale Method (HMM) \cite{multiscale2} which aims to minimize the computational cost by using MD simulation only in a fraction of the computational domain. Although clearly related to the approach discussed here, the two approaches are also significantly different. The work by Borg et al. is in principle more general (can be applied to high-aspect-ratio problems other than of lubrication type), but relies on online MD simulations; in contrast, we believe that a GLE approach would be preferable for treating lubrication problems for a number of reasons. First, beyond developing the closure that specifies the flowrate as a response to the pressure gradient, online MD simulations and numerical approximations (e.g. interpolation between MD domains such as the one used in \cite{borg}) are not required; in fact, this simplicity in some cases enables analytical solutions (as is the case for the validation problem considered here); moreover, the closure could also be obtained from experiments. Second,  
a GLE can be seamlessly integrated (both from a theoretical and a code development point of view) with other lubrication equation analyses (e.g. matching of solutions in different domains), but even more generally, with other analyses already developed to interface with lubrication equation treatments.  

Although a comprehensive review of previous work on models of transport beyond the Navier-Stokes is outside the scope of our work, we would like to briefly discuss some recent findings that are relevant to our MD results discussed in section \ref{sec:validation}. First we note that a number of recent studies suggest that deviations from the traditional no-slip Navier-Stokes description are due to the presence of additional effects that can be taken into account within the Navier-Stokes constitutive framework, rather than complete failure of the latter   \cite{exclusion-bocquet,Schlaich,extendCFD}. These effects include (of course) slip, different viscosities in different regions of the physical domain \cite{Schlaich} due to fluid layering \cite{JerryPRF}, a reduction in the effective domain size due to the gap between the solid and fluid \cite{exclusion-bocquet, BeskokJChemPhys} caused by the fluid-solid interaction \cite{Jerry, JerryPRF}, a reduction in the effective fluid density \cite{BeskokJChemPhys} due to fluid layering, and others. On the other hand, it is fair to say that no definite conclusion has been reached yet, since no general agreement exists on what these effects are and whether they are specific to the fluid-solid system under consideration or even the flow geometry.  Moreover, some studies find changes to the constitutive relation that are incompatible with the Navier-Stokes description \cite{evans97} or of unclear origin \cite{Beskok}. As a result, in the context of the Navier-Stokes description we will use the term "failure" (or the related expression "beyond Navier-Stokes") to refer to deviations from the {\it traditional} macroscale Navier-Stokes description, that given the above discussion, may not {\it necessarily} be a result of complete failure of the NS constitutive framework. In fact, as already pointed out before and extensively discussed below, the GLE is an attempt to enable solutions of a particular class of problems (lubrication) without requiring a resolution to the above questions.

The present paper is organized as follows. In the next section (section \ref{sec:method}), we briefly review lubrication theory and show how it can be generalized to arbitrary lengthscales. In section \ref{sec:validation}, we validate the GLE approach 
using an example lubricant-wall combination and compare the results against full MD simulation. We conclude
with section \ref{sec:conclusion}, in which we review our results and discuss possible extensions and improvements, as well as more general directions for future work.

\section{Extending the Reynolds Equation Beyond Navier-Stokes}
\label{sec:method}
The narrow-gap assumptions that underlie classical lubrication theory (see \cite{leal, cameron, szeri}  
for a detailed description) remain applicable to a large 
class of problems even at nanometer-size lengthscales \cite{gallis,garcia}. 
We start the development of a lubrication equation valid at those lengthscales by reviewing the main ingredients associated with the classical 
(Navier-Stokes-based) Reynolds equation.
\subsection{Background: The Reynolds Equation}
\label{subsec:background}
\begin{figure}
\centering
\begin{tikzpicture}[join=round]
    \tikzstyle{conefill} = [fill=blue,fill opacity=0.5]
    \tikzstyle{ann} = [inner sep=0.5pt]
    \tikzstyle{ghostfill} = [fill=white]
    \tikzstyle{ghostdraw} = [draw=blue!50]
	\coordinate (O) at (0.0,3); 
    \draw [cyan,name path = A,line width=3pt] plot [smooth] coordinates {(-5,6.5) (-2,4.5)
(1,5.5) (4,4.5) (6,5.5)}; 
  	\draw[arrows=<->,cyan,line width=3pt,name path = B](-5.5,3.0)--(5.5,3.0);
    \node[ann] at (5.65,3.0) {$\xcap$};
	\draw[arrows=<->,line width=1.5pt](0,1.5)--(0.0,6.5);
	\node[ann] at (0,6.75) {$\ycap$};
	\node[ann] at (0.2,2.75) {$\zcap$};
	\filldraw (0,3) circle (3pt); 
	\node[ann] at (1.5,4.25) {$h(x)$};
	\draw[arrows=.->,line width=1.5pt](1,3.0)--(1.0,5.5);	
	\node[ann] at (-2.0,2.65) {$U$};
	\draw[arrows=->,line width=1.5pt](-1.5,2.4)--(-2.5,2.4);
\end{tikzpicture}

		\caption{Sketch of a typical lubrication analysis geometry. }
		\label{fig:geo3}
\end{figure}
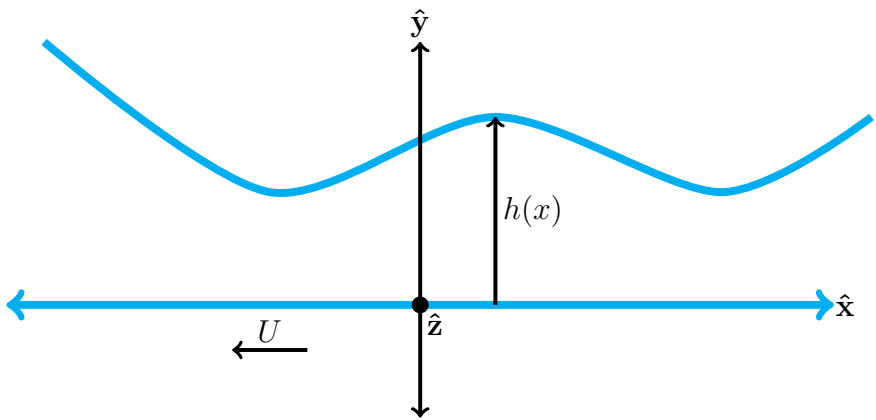

Consider an incompressible thin liquid film as shown in figure \ref{fig:geo3}. Let $x,y,z$ denote the cartesian coordinates along the $\xcap$, $\ycap$ and $\zcap$ directions, respectively, and  $u$, $v$, $w$ denote the components of the fluid velocity vector, $\bf u$, in the same respective directions. As shown in figure \ref{fig:geo3},  the film thickness is aligned with the $\ycap$ direction.  In the interest of simplicity, we will limit the discussion to flows of the type depicted in the  figure, in which no flow exists in the $\zcap$ direction (also $\partial/\partial z=0$) and only the lower boundary moves in the negative $\xcap$ direction with speed $U$, while the gap height is characterized by $h=h(x)$.

Let $h_0$ denote the characteristic film lengthscale in  the transverse direction, $\ycap$, and $L$ the characteristic lengthscale in the axial direction $\xcap$. Under the assumption $\varepsilon=h_0/L\ll 1$, the dynamical flow variables can be written in terms of their perturbation expansions in powers of $\varepsilon$; for example,  the flow velocity in the axial direction can be expanded as $u=u^0+\varepsilon u^1+O(\varepsilon^2)$. Considering only leading order terms and using impermeability constraints at the walls, 
the continuity equation can be integrated over the film thickness to obtain \cite{leal} 
\begin{align}
		\frac{\partial h }{\partial t} + \frac{\partial Q^0}{\partial x} &= 0
		\label{eqn:rey}
\end{align}
where 
\begin{equation}
    Q^0= \int_0^{h(x)} u^0 dy 
    \label{Q-equation}
\end{equation} 
    is the (local) volumetric flow rate. The above discussion follows the derivation in \cite{leal}; derivation of the Reynolds Equation in the presence of slip at the boundaries is discussed in \cite{burg}. 
    
    The traditional Reynolds lubrication equation 
\begin{align}
		\frac{\partial h}{\partial t} = \frac{\partial}{\partial x}\left( \frac{h^3}{12 \mu} \frac{dp^0}{dx} +   \frac{Uh}{2}\right).
		\label{eqn:basic-lubrication}
\end{align}
is obtained through the substitution
\begin{equation}
Q^0=-\frac{h^3}{12 \mu} \frac{dp^0}{dx}-\frac{Uh}{2}
\label{NSflowrate0}
\end{equation} 
in (\ref{eqn:rey}), where $\mu$ denotes the fluid viscosity and $p^0$ denotes the zeroth-order term in the expansion of the pressure field $p=p^0+\varepsilon p^1+O(\varepsilon^2)$, related to $u^0$ via the Navier-Stokes equation for fully-developed---not changing in the flow direction, due to, for example, entrance effects---pressure-driven flow 
\begin{align}
        \frac{d p^0}{d x} = 
		{\mu} \frac{\partial^2 u^0}{\partial y^2}
        \label{eqn:NSE}
\end{align}

In expression (\ref{NSflowrate0}), rewritten here 
\begin{equation}
Q=-\frac{h^3}{12 \mu} \frac{dp}{dx}-\frac{Uh}{2}
\label{NSflowrate}
\end{equation} 
without the superscript $0$, which will be henceforth omitted in the interest of simplicity,
the first term represents the pressure-driven (Poiseuille) component of the flow rate, while the second term represents the flow rate due to the motion of the lower boundary and will be referred to as the Couette component of the flowrate. They will be denoted by $Q_P^{\rm NS}$ and $Q_C^{\rm NS}$, respectively, where the superscript ${\rm NS}$ highlights their origin in the Navier-Stokes description (\ref{eqn:NSE}).

\subsection{Generalized Lubrication Equation}
\label{subsec:general}
We begin by noting that equation (\ref{eqn:rey}) is a statement of mass conservation and is thus valid at all lengthscales; the assumption of Navier-Stokes behavior only enters via (\ref{NSflowrate}). In other words, if an expression for $Q=Q_P+Q_C$ valid beyond Navier-Stokes can be developed, the resulting lubrication equation will also be valid beyond Navier-Stokes.   

We also note that, provided the upper and lower bounding surfaces are identical in structure and as far as wall-fluid interactions are concerned, the Couette 
flowrate remains the same as above, that is, $Q_C = Q_C^{\rm NS}$; in particular, 
if $-U \; \xcap$ is the relative velocity between the walls, $Q_C = -U h/2$. 

To introduce a more general form for the pressure-driven component of the flowrate we argue that it is reasonable to expect that the  fluid response {\it in fully developed flows} is proportional to the pressure gradient and thus can be written in the form
\begin{align}
\displaystyle Q_P = -\tilde{Q}_P(A,h,\rho,T,B) \frac{d p}{d x},
\label{eqn:model1}
\end{align}
where  $A$ denotes the (set of parameters characterizing the) fluid of interest, $B$ denotes the (set of parameters characterizing the) boundary interaction for the particular fluid-wall combination considered, $\rho$ denotes the fluid density and $T$ denotes the temperature. The more general form of the lubrication equation is thus
\begin{align}\label{general}
		\frac{\partial h}{\partial t} = 
		\frac{\partial }{\partial x}\Bigg( \tilde{Q}_P(A,h,\rho,T,B)\frac{dp}{dx}   + \frac{U h}{2}  
		\Bigg)
\end{align}

One might expect that  $\tilde{Q}_P \sim {\cal O}( h^n)$, with $1 \leq n \leq 3$, 
since $n = 3$ for Navier-Stokes with no-slip (and thus $\tilde{Q}_P\propto h^3$ as $h/\hat{\sigma} \rightarrow \infty$), while $n=1$ appears to be a reasonable lower bound, since the flow rate is expected to be proportional 
to the channel width. Recalling the dilute gas case \cite{fukui} is instructive: in this case, the pressure-driven flowrate can be written as 
\begin{equation}
    Q_P=-h^2 \sqrt{2RT}{\cal Q}({\rm Kn},B)\frac{dp}{dx} 
    \label{dilute-case}
\end{equation}
where ${\cal Q}({\rm Kn},B)$ is a dimensionless coefficient that depends on ${\rm Kn} =\lambda/h$, where $\lambda=m/(\sqrt{2}\pi \rho \sigma^2)$ is the mean free path, $\sigma$ is the hard-sphere diameter, $R$ is the specific gas constant and $B$ represents the effect of the gas-wall interaction (e.g. via one or more accommodation coefficients) \cite{cercig1}.  Therefore, 
$\tilde{Q}_P(A,h,\rho,T,B)=h^2\sqrt{2RT}{\cal Q}({\rm Kn},B)$, where $A=\{m,\sigma, R\}$.

Unfortunately, $v_m=\sqrt{2RT}$ is unlikely to be an appropriate characteristic molecular velocity for the dense-fluid case. In other words, $\tilde{Q}_P=h^2 v_m {\cal Q}(A,h,\rho,T,B)$ is unlikely to correctly scale dense-fluid flowrate data. Despite progress towards developing quantitative models of flow under extreme nanoconfinement (see \cite{Schlaich} for an example involving water), in the absence of a general description for the flowrate, $\tilde{Q}$ can be scaled using the traditional macroscopic formulation, namely $\tilde{Q}_P=h^3/(12 \mu(\rho,T)) F(A,h,\rho,T,B)$, where $F(A,h,\rho,T,B)$ is an {\it arbitrary} function to be determined by MD simulations or experiments for the fluid of interest and the requirement that  $F(A,h,\rho,T,B)\rightarrow 1$ as $h/\hat{\sigma}\rightarrow \infty$.

We close by noting that although some "empiricism" is required currently, the simplification introduced by the GLE is still considerable: as with any constitutive relation, $\tilde{Q}_P$ need only be determined once, in a fully developed lower-dimensional setting (from a plane Poiseuille flow). Once determined, equation \eqref{general} can be applied to arbitrary geometries (provided they satisfy the appropriate lubrication approximation criteria).

\section{Validation}
\label{sec:validation}
In this section we validate the ideas discussed in section \ref{subsec:general} by comparing solutions of the GLE for a particular fluid-solid combination  with MD simulations of the same system in a model lubrication  geometry. The working fluid in this validation problem is n-hexadecane, while the solid walls are iron. In the following subsection we use MD simulations for  determining $\tilde{Q}_P$ for the n-hexadecane-iron fluid-wall system. In subsection \ref{subsec:AnalyticalSolution} we obtain an analytical solution of (\ref{general}) for the model  problem based on the constitutive relation of section \ref{constitutive}. In subsection \ref{subsec:Comparison} we compare MD simulation results with the obtained analytical solution.

\subsection{Determining $\tilde{Q}_P$}
\label{constitutive}
In order to determine the unknown constitutive relation $\tilde{Q}_P(A,h,\rho,T,B)$ we perform MD simulations in 2-D channels under fully developed conditions for a wide range of $h$ values, namely $ 1 \; {\rm nm} \lesssim h\lesssim 11 \; {\rm nm}$, at a pressure of $p_0 = 80 \; {\rm MPa}$ and temperature of $T_0=450\;{\rm K}$. More details on the MD simulations can be found in Appendix A. 

Our validation problem was designed with typical lubrication problems in mind, in which the wall-fluid interaction does not vary as a function of space and thus no variation in $B$ need be considered. Moreover, we note that in typical nanoscale problems flow velocities are sufficiently small to make the isothermal assumption well-justified. Given the above, as explained in detail in the Appendix, care was taken to ensure that our validation problems were also consistent with these assumptions. Specifically, $U$ was sufficiently small for  temperature and density variations to be small (the latter due to incompressibility of the dense fluid \cite{szeri}), justifying the approximation  $\tilde{Q}_P(A,h,\rho,T,B)\approx \tilde{Q}_P(A,h,\rho_0,T_0,B_0)$. By verifying that shear rates in all our simulations were sufficiently small for linear response to be valid (see Appendix), we can thus expect variations in $\tilde{Q}_P$ to come only from variations in $h$, that is, we can proceed by taking $\tilde{Q}_P(A_0,h,\rho_0,T_0,B_0)=\tilde{Q}_P(h)$.

One possible form for  $\tilde{Q}_P(h)$ follows from using the fact that the flow rate must approach the well-known slip-corrected Poiseuille flowrate for $h/\hat{\sigma} \gg 1$. Specifically, we use the form 
\begin{align}
\tilde{Q}_P = \left(\frac{h^3}{12\mu}\Big(1+\frac{6L_s}{h}\Big)\right) \times F(h),
\label{eqn:model2}
\end{align}
where $L_s$ denotes a slip length, while $F(h)$ is  responsible for capturing deviations from the slip-corrected Poiseuille flowrate and thus needs to satisfy the requirement $F(h)\rightarrow 1$ as $h/\hat{\sigma}\gg 1$. Here it is important to note that the form (\ref{eqn:model2}) does not imply an underlying assumption of slip-flow, since $F(h)$ is still {\it arbitrary} (to be determined for each fluid from MD or experimental data). Writing (\ref{eqn:model2}) is akin to setting out to describe the dilute gas data with an expression of the form $\tilde{Q}_P=h^3(1+6L_S/h)\sqrt{2RT}{\cal Q}'/12 \mu$, which amounts to the rescaling ${\cal Q}'=12 \mu{\cal Q}/(h+6L_S)$. Clearly the two are equivalent and do not imply the presence of any restrictions on $\tilde{Q}_P$. 

\begin{figure}[ht!]
		\centering
		\includegraphics[width=1\textwidth]{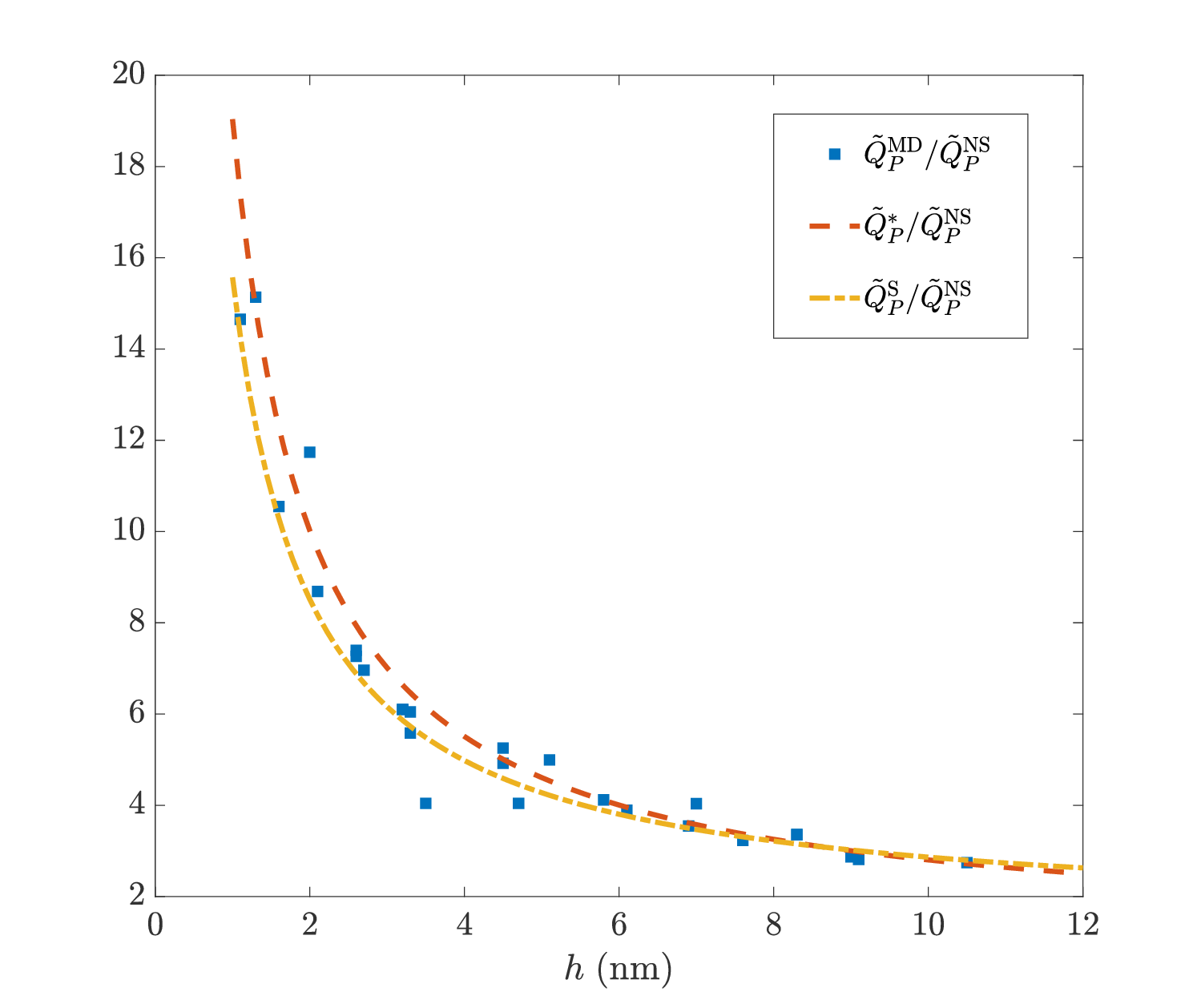}
		\caption{Flowrates for a range of channel heights $1.4 \; \text{nm} \; \leq h \leq 10.8 \; \text{nm}$. Comparison between the MD simulation results (denoted by $\tilde{Q}_P^{\rm MD}$) and expression (\ref{eqn:model3}) (denoted by $\tilde{Q}^*_P$). Expression (\ref{eqn:model2}) with $F=1$ and parameters  $\mu=\hat{\mu}$
and $L_s=\hat{L}_s$ (slip flow with parameters extracted from large channels) is also shown and is denoted by ${Q}^{\rm S}_P$. All results are normalized by  $\tilde{Q}_P^{\rm NS}$ (no-slip Navier-Stokes flowrate).}
		\label{fig:qbyqnsns}
\end{figure}

Our MD simulation results for the flowrate are summarized in figure \ref{fig:qbyqnsns}. Surprisingly, we find that choosing $F = 1 \; \forall \; h$ results in a good least squares fit to the flowrate.  In other words, for {\it the particular system studied here} the constitutive relation for the flowrate is given by 
\begin{align}
\tilde{Q}^*_P = \frac{h^3}{12\mu^*}\Big(1+\frac{6L_s^*}{h}\Big).
\label{eqn:model3}
\end{align}
where $\mu^* =0.67 \; {\rm mPa s}$ and $L_s^* =3.0 \; {\rm nm}$ have been determined by a least squares fit to the MD data shown in the figure.

This finding, namely that the MD flowrates can be described by a slip-corrected
Navier-Stokes expression (\ref{eqn:model3}) (i.e., that $F=1$ in (\ref{eqn:model2})) {\it for all} $h$ is a rather unexpected and perhaps fortuitous result, 
likely specific to the fluid-solid system considered here. We note that the values  $\mu^* =0.67 \; {\rm mPa s}$ and $L_s^* =3.0 \; {\rm nm}$, determined from a least-squares fit to all flowrate data ($1.4\;{\rm nm}\leq h\leq 10.8\;{\rm nm}$), are quite different from the viscosity and slip length obtained by fitting the slip-corrected Poiseuille velocity profile to our $y$-resolved MD data for $h\gtrsim 6\; $nm  where the slip-flow description is expected to be valid. 

\begin{figure}[ht!]
		\centering
		\includegraphics[width=\textwidth]{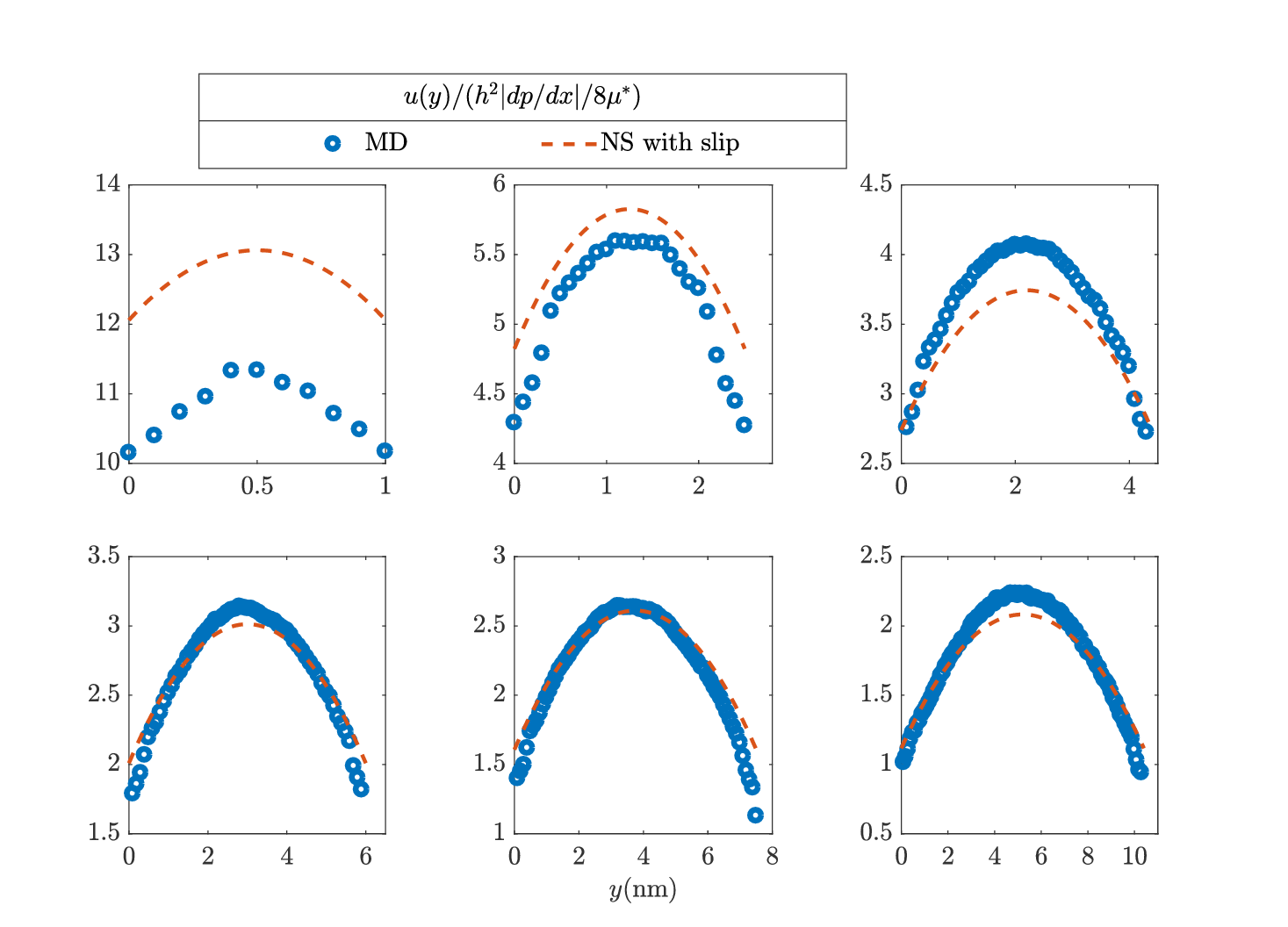}
		\caption{Normalized velocity profiles for films of different
thicknesses in "pressure-driven" flow. Blue circles denote MD simulation data, while the red lines 
are the velocities predicted by slip-corrected Navier-Stokes with parameters
$L_s^*$ and $\mu^*$. The channel widths considered in the figures
$A$ to $F$ respectively are $1.4, 2.9, 4.8, 6.4, 7.8$ and $10.8\; {\rm nm}$.}  	\label{fig:ubyrhog}
\end{figure}

Selected $y$-resolved velocity profiles are shown in figure \ref{fig:ubyrhog}. This figure shows that the coefficients $\mu^* =0.67 \; {\rm mPa s}$ and $L_s^* =3.0 \; {\rm nm}$, are not necessarily able to describe the $y$-resolved flow profiles in all cases. In particular, we find that for "large" $h$  (specifically, $h\gtrsim 6$ nm), individual $y$-resolved profiles are well described by a slip-corrected Poiseuille profile with parameters $\hat{\mu}=0.46\; {\rm mPa s}$
and $\hat{L}_s=1.6 \; {\rm nm}$.  In the range $3 \;\text{nm}\lesssim h\lesssim 6 \; \text{nm}$, although the velocity profiles appear parabolic, the coefficient of viscosity and slip length extracted from fits of $y$-resolved MD data to slip-corrected Poiseuille profiles are film-thickness ($h$) dependent; this is consistent with previous reports \cite{Beskok}. For $h\lesssim 3 \; {\rm nm}$, the velocity profiles are not parabolic. 
In other words, $\mu^* =0.67 \; {\rm mPa s}$ and $L_s^* = 3 \; {\rm nm}$ {\it must be interpreted as parameter values that reproduce  the flowrate data} when used in (\ref{eqn:model3}) but  {\it do not correspond to viscosity and slip length values in the Navier-Stokes sense}. The relatively small difference between $\tilde{Q}_P^*$ and the slip flowrate $\tilde{Q}^{\rm S}_P$ (obtained using parameters $\hat{\mu}$ and $\hat{L}_s$) shown in figure \ref{fig:qbyqnsns}, is also consistent with the findings of \cite{Beskok} which suggest that the flowrate may exhibit smaller devations from macroscopic results than other quantities, because as an integral quantity it is more sensitive to cancellation of competing effects.

\subsection{The barrel drop problem} 
\label{subsec:AnalyticalSolution}
\begin{figure}
\begin{tikzpicture}[join=round]
    \tikzstyle{conefill} = [fill=blue,fill opacity=0.5]
    \tikzstyle{ann} = [inner sep=0.5pt]
 	\node[inner sep=0pt] (russell) at (0,0)
    {\includegraphics[width=1\textwidth]{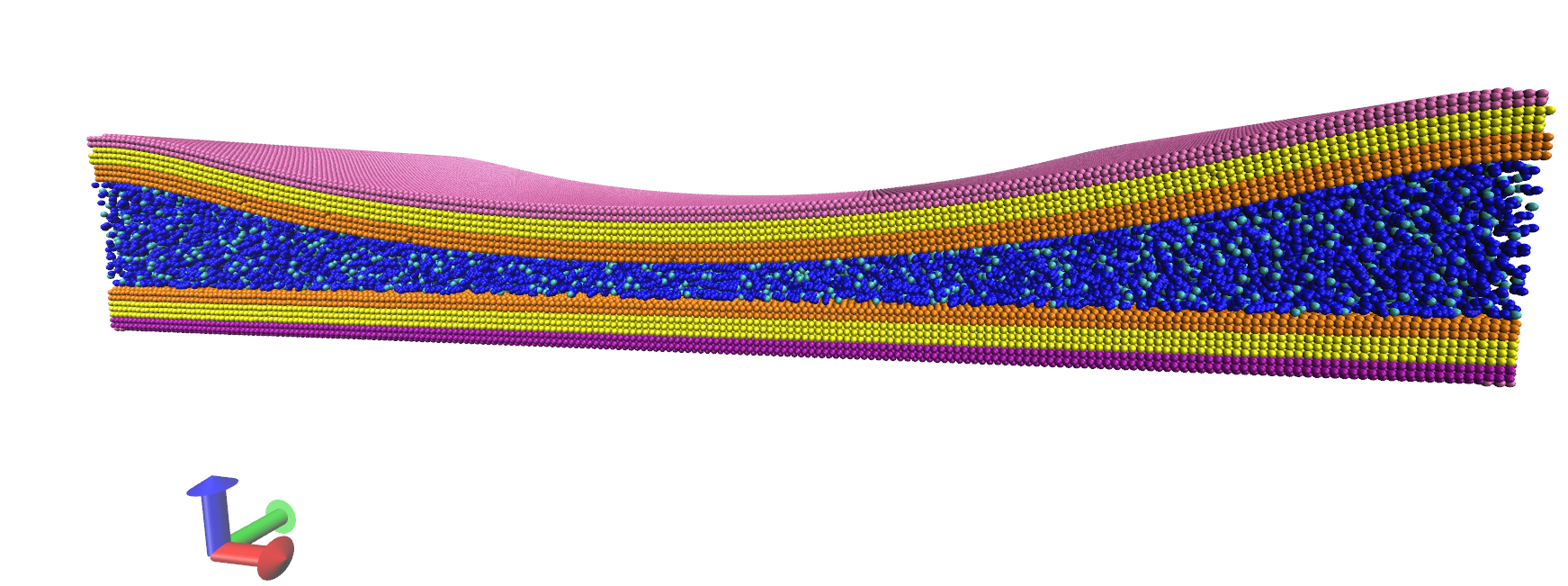}};
	\node[ann] at (-4.3,-2.6) {$\xcap$};
	\node[ann] at (-5.5,-1.4) {$\ycap$};
	\node[ann] at (-4.5,-1.85) {$-\zcap$};
	\draw[arrows=-,line width=2.5pt](-6.15,0.1)--(6.6,-0.2);
	\draw[arrows=->,line width=2.5pt](-6.15,0.1)--(-5.5,0.4);
	\node[ann] at (-6.5,0.1) {$A$} ;
	\draw[arrows=<-,line width=2.5pt](7.0,0.1)--(6.6,-0.2);
	\node[ann] at (7.25,-0.2) {$A'$};
	\filldraw (6.4,-0.75) circle (2pt); 
	\node[ann] at (7.25,-0.75) {$L_x$};
	\filldraw (-5.85,-0.35) circle (2pt); 
	\node[ann] at (-6.5,-0.75) {$0$};

		\node[ann] at (-2.0,1.7) {$h(x) = h_0 + a (x-L_x/2)^2$};
\end{tikzpicture}
\caption{
A schematic model of the "barrel drop" 
		lubrication geometry, which is used in the MD validation problem. $L_x$ is the size of the MD simulation box in the axial  
direction $\xcap$. The section at $A$-$A'$ shows the $x$-$z$ plane along
which the pressure is measured using the method of planes \cite{methodofplanes1} (See
\ref{app:mop}). }

\label{fig:channelGeometry}
\end{figure}

Figure \ref{fig:channelGeometry} shows the "barrel drop" geometry used for validation of the ideas presented in this paper. In this problem, flow occurs due to the motion of the lower wall in the negative $\xcap$ direction with velocity $U$, as assumed in the lubrication equation (\ref{general}). The upper wall is stationary and of parabolic shape. Thus, the gap height as a function of the axial coordinate is given by $h(x)=h_0+a (x-L_x/2)^2$, where $L_x$ is the 
length of the domain in the $\xcap$ direction.

In a general case, equation \eqref{general} would have to be solved numerically (say, using a Finite
Volume scheme). In this case however, the simplicity of the barrel drop geometry and the constitutive relation (\ref{eqn:model2}) 
makes a semi-analytical solution for the pressure distribution possible. Using the fact that $\partial h/\partial t=0$, we integrate \eqref{general} once to obtain the constant volumetric flowrate  
\begin{align}
		Q= -\frac{U h(x)}{2} - \frac{dp}{dx} \tilde{Q}^*_P(h(x)) 
\end{align}
where $\tilde{Q}^*_P(h(x))$ is given in (\ref{eqn:model3}). 
Solving for the pressure, we obtain
\begin{align}
		p(x) = p_0  - \int_{0}^{x}  \frac{  2Q + Uh(x')}{\displaystyle \tilde{Q}^*_P(h(x'))} \; dx'
		\label{eqn:analyticalPressure}
\end{align}
The flowrate $Q$ can be calculated from the following expression, 
\begin{align}
		Q = - \dfrac{ {\displaystyle  \int_{0}^{L_x}  } Uh(x')[\tilde{Q}^*_P(h(x'))]^{-1} dx' }{ \displaystyle{ \int_{0}^{L_x} } 
		2[\tilde{Q}^*_P(h(x'))]^{-1} dx' }
\label{eqn:periodicbc}
\end{align}
obtained
on applying periodic boundary conditions $p(0)
= p(L_x)$ on the pressure.
   The constant $p_0$ can be calculated from the force balance 
   \begin{equation}
       \int_{0}^{L_x} p(x) dx=N
       \end{equation}
       where $N$ denotes the total normal force per unit depth (in the $z$ direction) acting on the top surface (assumed given).
   Due to the symmetry of the film thickness profile about $x=0$, the above relation 
reduces to \cite{nishathesis}:
\begin{align}
p_0 L_x=  N
\label{eqn:integralbc}
\end{align}
 Finally, the analytical pressure distribution can be obtained from (\ref{eqn:analyticalPressure}), using $Q$ and $p_0$ from (\ref{eqn:periodicbc}) and (\ref{eqn:integralbc}).

\subsection{Results}
\label{subsec:Comparison}
\begin{figure}[ht!]
		\includegraphics[width=\textwidth]{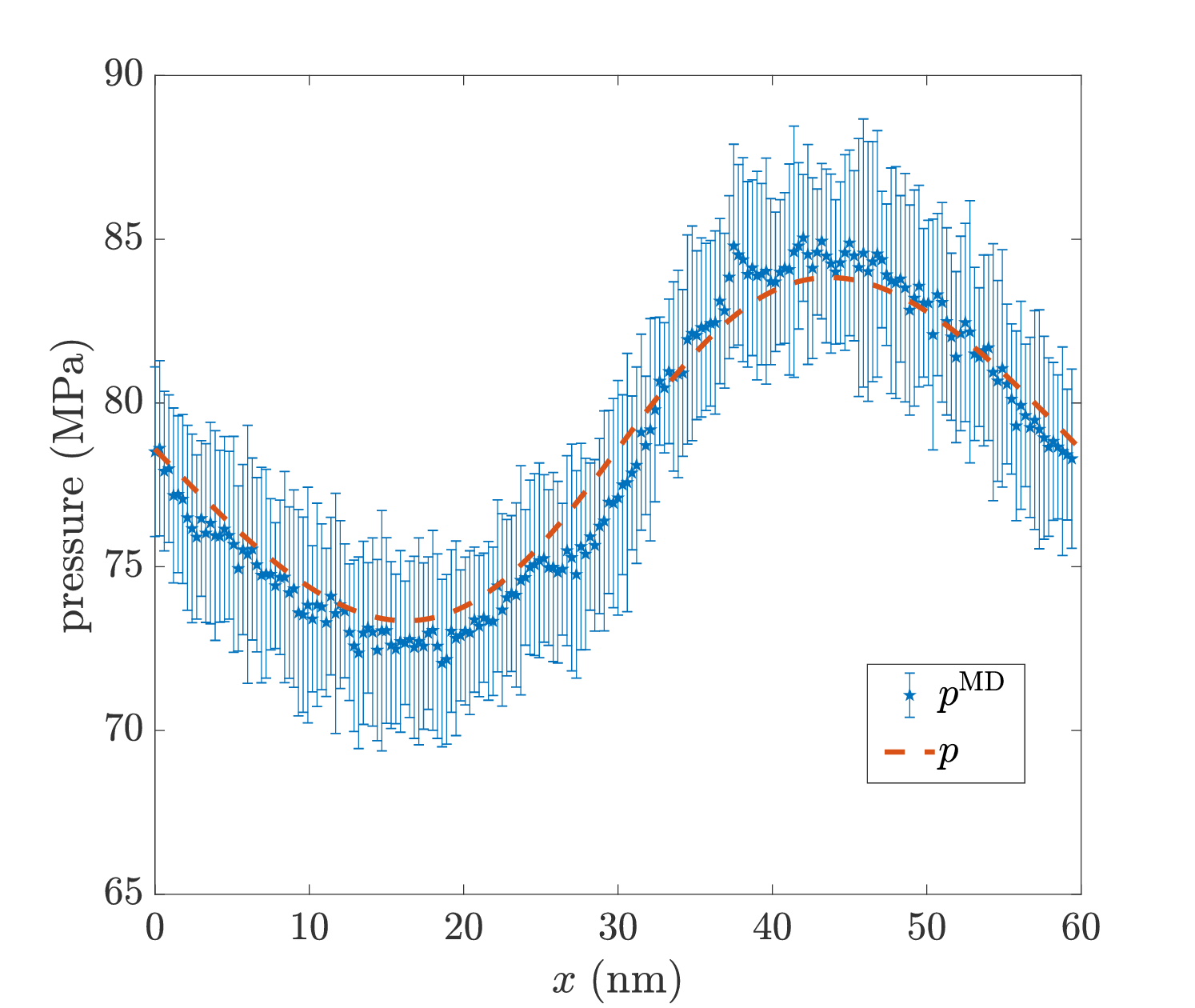}	
		\caption{Comparison between the
				semi-analytical solution $p(x)$ given by (\ref{eqn:analyticalPressure}) and 
the pressure computed from the MD simulation, $p^{\rm MD}(x)$, for the "barrel drop" problem with $h_0=3.8 \; {\rm nm}$. }		 
		\label{fig:compp}
\end{figure}

In this section we compare the pressure distribution 
\eqref{eqn:analyticalPressure} 
against the corresponding MD result, denoted by $p^{\rm MD}$. Two such comparisons are performed here for different 
minimum gap heights ($h_0$). The first one 
is for a minimum gap height $h_0=3.8\; {\rm nm}$, while the second one is for $h_0=1.6 \; {\rm nm}$. In the first comparison, shown in figure \ref{fig:compp}, $h_0$ is sufficiently large to avoid the presence of solvation and disjoining effects which make direct comparison 
of the pressure fields difficult. The pressure computed from the MD simulation
agrees with the analytical pressure distribution closely. 
Here we recall that, according to our MD results, for $h\lesssim 6  \; 
{\rm nm}$, $y$-resolved velocity profiles cannot be described by a fixed ($h$-independent) viscosity and slip-length, and thus a lubrication approach based on the traditional Reynolds equation would not be valid.

As stated above, validation at smaller gap heights is complicated by 
the appearance of solvation (structuring) and disjoining pressure effects \cite{israelachvili,kato}.  Disjoining pressure effects are caused by long-range (van der Waals) forces and have been considered in a variety of contexts, most notably wetting \cite{degennes}. Solvation forces result from the entropic contribution of fluid layering at the wall \cite{israelachvili};  as a result, they are important when the characteristic lengthscale $h$ is comparable to the layering thickness, that is, only under extreme confinement \cite{kato,joly}.   

\begin{figure}[ht!]
		\includegraphics[width=1\textwidth]{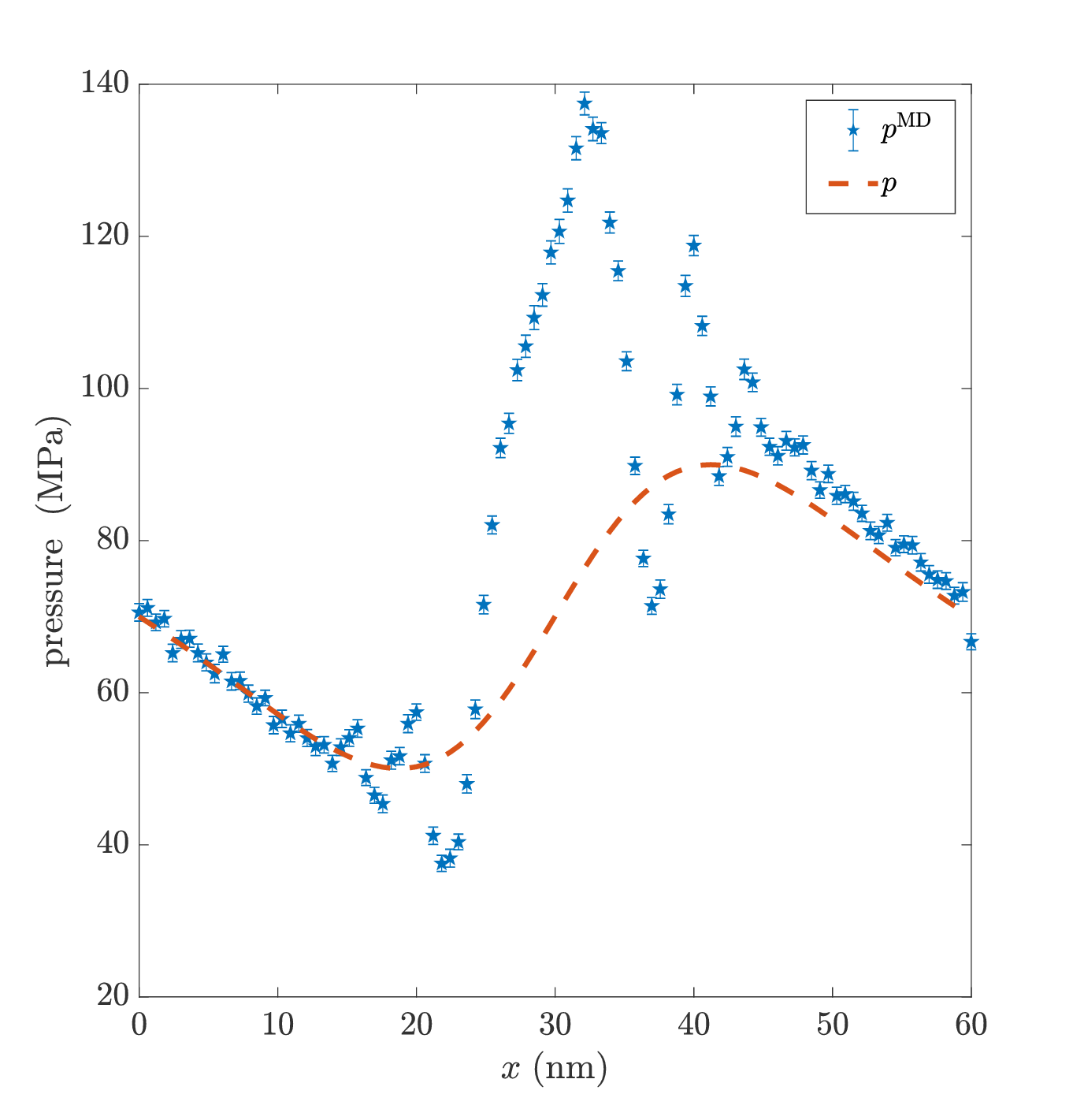}	
		\caption{Comparison between the
				semi-analytical solution $p(x)$ given by (\ref{eqn:analyticalPressure}) and 
the pressure computed from the MD simulation, $p^{\rm MD}(x)$, for the "barrel drop" problem with $h_0=1.6 \; {\rm nm}$  As $h\rightarrow \hat{\sigma}$, disjoining and solvation effects become important. The latter are particularly apparent. }		 
		\label{fig:pcomp_1nm_without_solvation}
\end{figure}

Figure \ref{fig:pcomp_1nm_without_solvation}  shows $p^{\rm MD}$ for the barrel drop case with $h_0=1.6\; {\rm nm}$; although both solvation and disjoining effects are important at this scale, solvation effects are particularly evident in the regions corresponding to $ h\lesssim 3\; {\rm nm}$ due to their distinctive oscillatory nature.   

\begin{figure}
		\includegraphics[width=\textwidth]{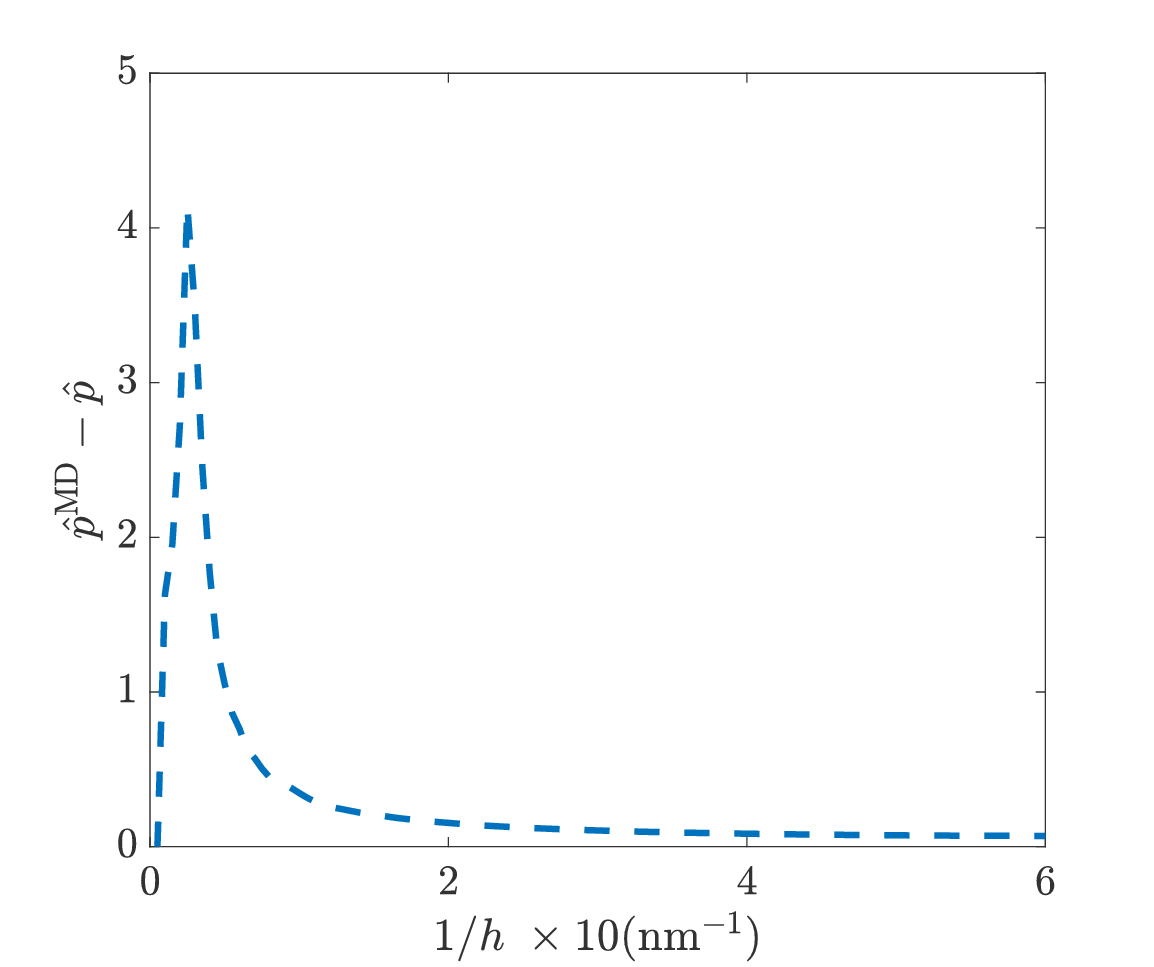}	
		\caption{The Fourier transform of $p^{\rm MD} -
p$ clearly shows a peak at $1/h = 2.5$ (corresponding to $0.4 \; {\rm nm}$), 
indicating the dominant wavelength associated with the spatial variation of solvation effects.}		 
		\label{fig:fft}
\end{figure}

Solvation and disjoining pressure effects can be included in the proposed formulation by solving equation (\ref{general}) subject to the boundary conditions   
$p(0) = p(L_x)$ and 
\begin{equation}
    N=\int_{0}^{L_x} [p(x)+p_s(h(x))+p_d(h(x))] \; dx
    \label{modifiedbalance}
\end{equation}
where $p_s(h(x))$ denotes the solvation pressure and 
\begin{equation}
    p_d(h(x))=\frac{A}{6 \pi h^3(x)}
\end{equation}
is the disjoining pressure, where $A$ denotes the Hamaker constant \cite{israelachvili}. 

\begin{figure}
		\includegraphics[width=\textwidth]{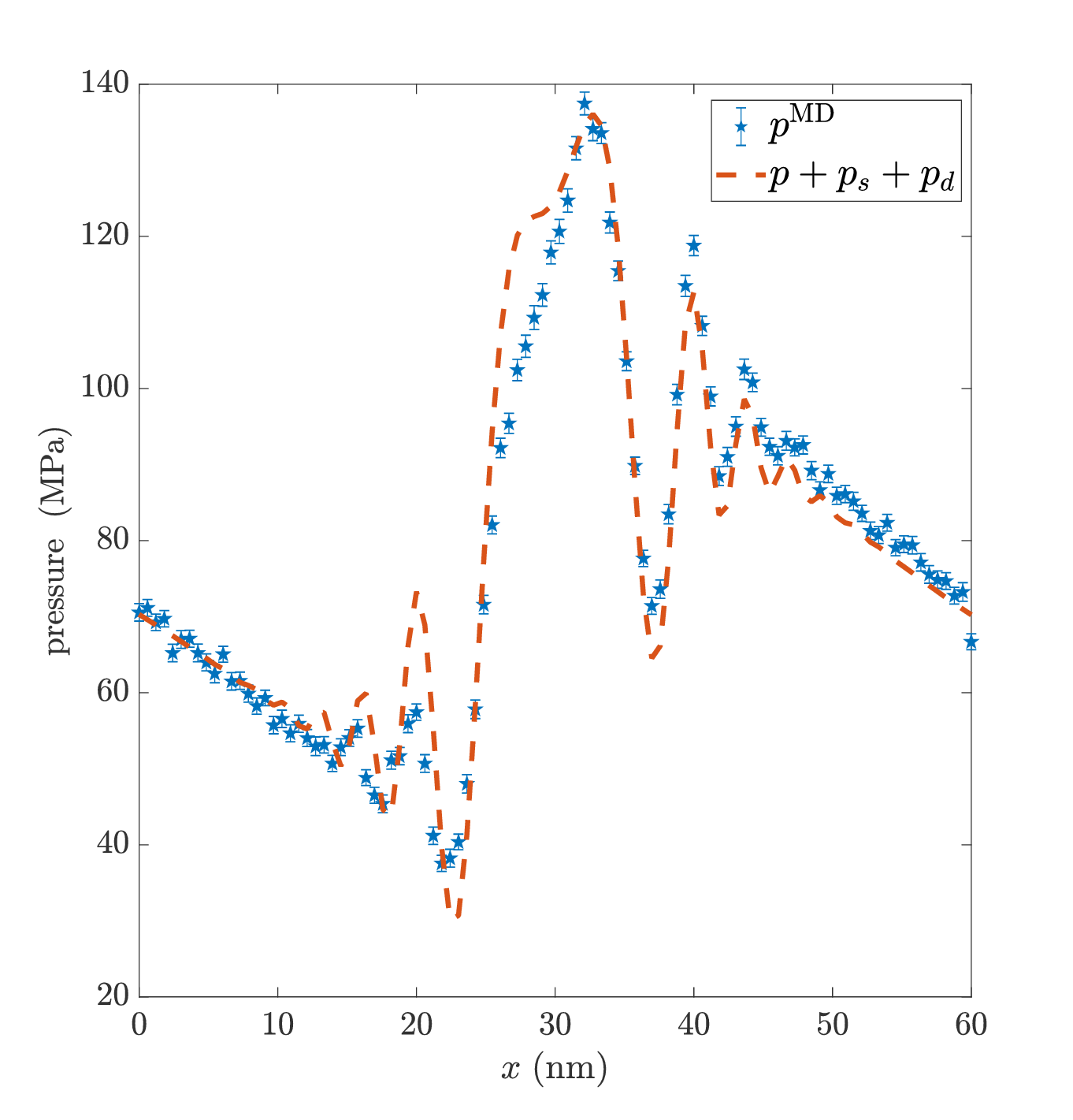}	
		\caption{Comparison between $p(x)+p_s(h(x))+p_d(h(x))$, given by (\ref{eqn:analyticalPressure}) and (\ref{eqn:Solvation}) respectively, and 
the pressure computed from the MD simulation, $p^{\rm MD}(x)$, for the "barrel drop" problem with $h_0=1.6 \; {\rm nm}$. }		 
		\label{fig:compp_1nm}
\end{figure}

It is important to note that although $p_s$ and $p_d$ are of $O(\varepsilon^0)$, they are not included in the GLE (\ref{general}) because, as discussed in section \ref{subsec:background}, the (local) flowrate $Q$ is calculated from the response to the zeroth-order balance (\ref{eqn:NSE}), while   
\begin{equation}
    \frac{d}{dx}(p_d+p_s)=\frac{d(p_s+p_d)}{dh} \frac{dh}{dx}
\end{equation}
is of order $\varepsilon^1$. 

 Although the molecular origin of the solvation pressure is well-understood \cite{israelachvili,henderson}, reliable expressions for predicting its magnitude for various fluids are not available. However, Joly et al. \cite{joly} have recently shown that an expression of the form 
\begin{align} 
p_s^{2D}(h(x)) &= -\rho_\infty k_B T \cos{(2 \pi h(x)/\delta)} 
							\exp{(-h(x)/\delta)}.
\label{eqn:Solvation}
\end{align}
developed for a hard-sphere fluid in two-dimensions, can be used to describe this effect in water, provided $\rho_\infty$, originally defined as the bulk fluid density far from the wall, was treated as an adjustable parameter. 
In the above expression, $\delta$ is a parameter related to the fluid structure and is thus expected to be of order $\hat{\sigma}$. We use $p_s^{\rm 2D}$ above to  model $p_s$, with 
the value of $\delta$ and $\rho_\infty$ determined from our simulation data. 
To determine $\delta$ we use the Fourier transform of $p^{\rm MD}-p$, shown in figure \ref{fig:fft}, which exhibits a single peak at
$0.4 \;\rm{nm}$, confirming the sinusoidal dependence of $p_s$ on $h$ and suggesting $\delta = 0.4 \; {\rm nm}$. This value is known from surface-force 
measurements and MD simulations \cite{ribarsky,sfa} to be close to the
 thickness of 
a layer of n-hexadecane confined between crystalline substrates.

Using this value for $\delta$ and treating $\rho_\infty$ as an adjustable (fitting) constant, we show in figure \ref{fig:compp_1nm} that the decomposition $p+p_s+p_d$ can model $p^{\rm MD}$ for $h$ as low as $1.6\; {\rm nm}$.

\section{Discussion and Conclusions}
\label{sec:conclusion}
We have discussed a Generalized Lubrication Equation framework for the solution of dense-fluid lubrication problems at the nanometer scale. The framework is based on the observation that under the lubrication approximation, the closure required is significantly less difficult to obtain than for the general flow case. 

 In the example considered here, it was assumed that the two solid surfaces interact with the fluid identically and thus the Couette component of the flowrate, by symmetry, remains equal to its Navier-Stokes no-slip value. The proposed methodology can be extended to cases where interactions are asymmetric by developing a constitutive relation for the Couette component of the flowrate.  Although in the example lubricant-wall model studied here the flowrate was reproduced using a slip-flow-based relationship, the GLE is in no way reliant on the validity of a  slip-flow description.  As an example, see the dilute gas case  (eq. \eqref{dilute-case}) and the related  discussion in section \ref{subsec:general}. Developing a {\it general} scaling relation for $\tilde{Q}_P$ in terms of the governing molecular parameters for dense fluids is the subject of future work, although significant progress has been made for particular fluid-solid systems (see section \ref{intro}). Such a relation would potentially reduce the number of simulations/experiments required to describe $\tilde{Q}_P(A,h,\rho,T,B)$ for a particular fluid.

 Developing a constitutive relation and solving the resulting GLE is considerably more tractable than brute-force MD simulations for {\it problems of practical interest}. In fact, the latter is usually impossible, both because of the cost associated with the number of particles required for such simulations, but also because the characteristic timescales of such systems are much longer. For example, the cost associated with the GLE  (primarily MD simulations for constructing the constitutive relation) is smaller (approximately by a factor of 2) than the cost of one  barrel drop simulation, even though the latter is at lengthscales much smaller than actual problems of practical interest (deliberately chosen for validation purposes, to ensure that an MD solution was feasible). Once the constitutive relation is developed, additional solutions of the GLE come at a cost that is essentially negligible (compared to MD simulations).  
   
For very small film heights, solvation and disjoining effects become important. Fortunately, as in previous work \cite{joly,kato}, it was shown that these effects can be satisfactorily accounted
for by linearly superposing their contribution to the hydrodynamic pressure, requiring minimal modification of the lubrication equation. Although in both cases (present and previous work \cite{joly}) good agreement with MD simulations was found by fitting the result for hard-spheres in two dimensions, it is clear that our ability to model solvation effects for more realistic fluids is very limited and needs to be improved.

\section{Acknowledgments}
The authors would like to thank Mathew Swisher and Gerald J. Wang for help with the computations and many helpful comments and suggestions during the course of this research. This work was sponsored by the Consortium on Lubrication in Internal Combustion Engines with additional support from Argonne National Laboratory and the US Department of Energy. The current consortium members are Daimler, Mahle, MTU Friedrichshafen, PSA Peugeot Citro\"en, Renault, Shell, Toyota, Volkswagen, Volvo Cars, Volvo Trucks, and Weichai Power. The use of the Center for Nanoscale Materials, 
an Office of Science user facility, was supported by the U. S. Department of Energy, 
Office of Science, Office of Basic Energy Sciences, under Contract No. DE-AC02-06CH11357.

This is a pre-print of an article published in Microfluidics and Nanofluidics. 

\appendix
\section{MD simulations}
\label{sec:mddetails}

 All MD simulations in this work were performed using the 
  software package LAMMPS \cite{steve}.

As explained in section \ref{subsec:general}, the constitutive relation was developed by studying 
fully-developed flow in two-dimensional channels. The validation was performed in the barrel drop geometry of figure \ref{fig:channelGeometry}, which is a significantly larger scale problem. In both cases,  the model lubricant is n-hexadecane, simulated using the TraPPE potential \cite{martin}, while the system solid boundaries are 
atomically smooth iron (BCC Fe) walls, simulated using the Embedded Atom
Method - Finnis Sinclair \cite{nist,eam,ackland} potential. The 
Lennard-Jones (LJ) parameters used between well-separated pseudo-atoms, i.e, 
a pair of  
pseudo-atoms separated by 3 or more pseudo-atoms within the same molecule
or a pair belonging to two different molecules within
the fluid are given in table \ref{table:params}.
The LJ parameters of interaction between atoms of unlike types were calculated using
the Lorentz-Berthelot mixing rules.

Bond stretching in the fluid molecules was modelled via a harmonic potential
of the form, $\displaystyle U_{ij} = \frac{k_r}{2} (r_{ij} - r_0)^2$, where 
$r_0 = 1.54 \; \angstrom $  is the equilibrium bond length and 
$k_r$, the bond stretching parameter taken here to have the value  $19.5139\; {\rm eV/\angstrom}^2$ 
(chosen from Lopez-Lemus et al \cite{lopez}).
The bending energy associated with variations of the angle between 3 adjacent atoms connected through 
bonds was modelled through a harmonic potential \cite{martin} of the form 
$\displaystyle U_{{\rm bend}} =  \frac{k_\theta}{2} (\theta - \theta_0)^2$, where
the bending parameter $k_\theta$ is taken as $2.6925\; {\rm eV}$ and
the equilibrium angle, $\theta_0 = 114^\circ$. 
The torsional potential that arises due to the bending of the 
dihedral angle, $\phi$, was given by an OPLS style potential \cite{watkins} :
$\displaystyle U_{{\rm torsion}} = \frac{c_1}{2}(1 + \cos \phi) 
+ \frac{c_2}{2} ( 1 - \cos(2 \phi)) + 
\frac{c_3}{2} ( 1 + \cos(3 \phi)) $.
The parameters were taken from Martin et al. \cite{martin} to be:
$$ c_1 = 0.05774 \quad c_2 = - 0.0117524, \quad c_3 = 0.136382 $$
\begin{table}
\centering
\begin{tabular}{|l | c | c | r|}
		\hline
		Type of pseudo-atom $i$ & Type of pseudo-atom $j$ & 
		$\sigma$(\AA) & $\epsilon$(eV) \\
		\hline
		$CH_2$  & $CH_2$ & 3.95 & 0.003964 \\
		$CH_3$ & $CH_3$ & 3.75 & 0.008445 \\
		$CH_3$ & $CH_2$ & 3.85 & 0.0058 \\
		$Fe$ & Fe & 2.2 &  0.02947 \\
		\hline
\end{tabular}
		\caption{The LJ parameters of interaction between different atomic species used 
		in the paper.}
		\label{table:params}
\end{table}

The interactions between the solid and the liquid atoms 
were modeled the LJ potential. The parameters 
for the wall-fluid interaction potential were computed 
through the Lorentz-Berthelot mixing rules, using the 
self-interaction parameters listed in table \ref{table:params}.
Note that the self-interaction parameters for Fe were taken 
from Berro et al \cite{berro} (developed for more realistic
simulation of Fe oxide surfaces) and are different
from those reported in the literature for BCC Fe (see 
Zhen et al. \cite{zhen}, for example). 

\subsection{Channel flow}
The simulations were performed in channels of length $99.94 \; \angstrom$  in the flow direction, $\xcap$, and width $71.38 \; \angstrom$, in the $\zcap$ direction. The channel width in the transverse $\ycap$ direction, denoted by $h$,  varied between $1.4 \; {\rm nm}$ and $10.8 \;{\rm nm}$. Fluid motion was generated by 
applying a body force per unit mass in the flow direction, $g$, while periodic boundary conditions were imposed in the flow direction, $\xcap$, thus eliminating entrance effects and making the flow fully developed by construction. Periodic boundary conditions were also applied in  the homogeneous (depth) direction, $\zcap$. In the transverse, $\ycap$, direction, the fluid was bounded by two solid walls.   
Each wall consisted of eight layers of
atomically smooth BCC Fe (001) surface, of thickness $ 22.84 \;\angstrom$.     The two layers furthest away from the fluid were frozen (using \verb+LAMMPS+'s 
\verb+fix rigid+)  and provided the wall structure; the adjacent three layers of atoms were responsible for maintaining the simulation temperature at 450 K and were thus thermostatted using a Nos\'e-Hoover thermostat.
No constraints were imposed on the three layers closest to the liquid. The fluid pressure was maintained at 80 MPa, by applying a normal force (in the negative $\ycap$ direction) to 
the frozen solid layers. To avoid thermostat-induced artifacts \cite{pahlavan}, no thermostat was applied to the fluid molecules during the data sampling phase (a Langevin thermostat was used for the initial equilibration phase). Viscous heat generated  within the fluid was removed by the thermostatted wall molecules. To ensure small temperature variations, the maximum fluid velocity was on the order of 50 m/s or less, resulting in a maximum temperature variation of $O$(5K), which is sufficiently small for the isothermal assumption to be valid.   We have determined empirically that taking the body force to be sufficiently small so that temperature variations were small (and thus non-linearities due to temperature-dependent transport coefficients were negligible) also ensured linear response, for which pressure-driven and body-force-driven flow are expected to be equivalent. This equivalence was validated by our results of section \ref{constitutive}. We also note that actual flow speeds encountered in nanoscale flows are much smaller (usually by many orders of magnitude) than the flow velocities used here (which are large in order to improve signal to noise ratio \cite{nicolas_error}); in other words, isothermal flow and linear response assumptions are even more justified for real flows.

Simulation results are shown in figures \ref{fig:qbyqnsns} and \ref{fig:ubyrhog} and discussed in section \ref{constitutive}. In figure \ref{fig:ubyrhog}, $|dp/dx|$ represents the "effective pressure gradient magnitude" ($=\rho g$).

\subsection{Barrel drop}

The geometry of the barrel drop profile is shown in figure \ref{fig:channelGeometry}. Fluid flow was generated by 
the motion of the lower solid boundary at a speed of $U = 60\;{\rm m/s}$. Both the upper parabolic surface and the lower plane surface consisted  
of rigid, thermostat and deformable layers, with constraints (including thermostating) analogous to those discussed in the previous section on channel flow. Specifically, heat was removed from the system by thermostating the three wall layers furthest away from the fluid in each wall, and no thermostat was applied to the fluid during the data sampling phase. The extent of the 
simulation box in the two lateral directions, $\xcap$ and $\zcap$, was $L_x = 59.9 \;{\rm nm}$ 
and $L_z = 25.6 \;{\rm nm}$, respectively. 
Based on a characteristic lengthscale $L_c =40 \; {\rm nm}$, the non-dimensional value of 
the curvature parameter is $aL_c^2/h_0 = 1.845$.

\subsection{Computation of pressure in MD simulations}
\label{app:mop}
As is well known from previous work \cite{methodofplanes1, methodofplanes2,methodofplanes3}, 
the Irving-Kirkwood (IK) formula for the stress tensor  in a fluid exhibits spurious oscillations near hard system boundaries such as solid walls. As a result, in the present work the fluid pressure is measured by calculating the $p_{yy}$ component of the configurational part of the stress tensor, as the sum of all molecular forces acting across a flat (imaginary) dividing plane whose normal is in the $\ycap$ direction.  Specifically, we are using an approach known as the Method of Planes (MOP) \cite{methodofplanes2}, in which the configurational part of the stress tensor at location $y$ is calculated by  
\begin{align}\label{stresstensor}
		p_{yy}^c(y) &= \frac{1}{4A}
\sum_{ij} F_{ij}^y \Bigg(
{\rm sgn}(y_i - y) - 
{\rm sgn}(y_j - y) \Bigg)
\end{align} 
where  
$y_i$ and $y_j$ are the components of the position
vectors of atoms $i$ and $j$ in the $\ycap$ direction at time $t$, respectively, and
$F_{ij}^y$ represents the $\ycap$-component of the 
force on atom $i$ due to atom $j$. The force component
 $F_{ij}^y$ contributes to the configurational pressure
only if it acts across the plane at $y$. The pressure is then given by $p= p_{yy}$;  we have independently verified, via simulations in 2D channels that the stress tensor remains isotropic even at these small scales.
The above definition assumes pairwise forces 
between contributing atoms. Stress tensor definitions for systems of particles interacting through many-body forces have only been developed for the virial IK-based formulation \cite{aidan}. To overcome this limitation, here we use the fact that, in the lubrication  approximation, the fluid pressure is independent of $y$. We have thus placed the sampling plane in the narrow gap between wall and liquid, which ensures that  any pair of
contributing atoms is never both solid or liquid. An additional benefit associated with this choice is that no atom-crossings occur across this sampling plane, meaning that the kinetic term 
\begin{align}
p_{yy}^k(y) &=  \frac{1}{2A}
\sum_{i} m_i v_i 
\frac{d}{dt} {\rm sgn}(y_i - y).
\end{align}
does not contribute. In the above equation, 
$m_i$ is the mass of particle $i$ and $v_i$ is the component of the particle velocity of
particle $i$ in the $\ycap$ direction at time $t$. 
 
Although derived for infinitely large planes,  in the case of the barrel drop geometry, in order to resolve the pressure variation in the flow direction, we have applied expression (\ref{stresstensor})
 locally, that is, over bins of finite extent in the flow ($\xcap$) direction. The number of bins was chosen to balance the need for good axial resolution of the pressure with the need for the number of bins to be as small as possible so that the change in pressure across consecutive bins is small.  Specifically, we used 100 bins, with 
each bin spanning $0.6\; {\rm nm}$ in the $\xcap$ direction and the size of the periodic box,
$25.6\; {\rm nm}$, in the $\zcap$ direction.  


\bibliographystyle{model1-num-names}
\bibliography{main.bib}

\end{document}